\documentclass{ws-procs10x7}
\newcommand{\etal}{{\it et al.}}
\usepackage{epsfig}
\begin{document}
\title{Charm Decays Within the Standard Model and Beyond}
\author{Marina Artuso}
\address{Department of Physics, Syracuse University, Syracuse NY 13244,
USA\\E-mail: artuso@phy.syr.edu}

\twocolumn[\maketitle\abstract{ The charm quark has unique
properties that make it a very important probe of many facets of
the Standard Model.  New experimental information on charm decays
is becoming available from dedicated experiments at charm
factories, and through charm physics programs at the b-factories
and hadron machines. In parallel, theorists are working on matrix
element calculations based on unquenched lattice QCD, that can be
validated by experimental measurements and affect our ultimate
knowledge of the quark mixing parameters. Recent predictions are
compared with corresponding experimental data and good agreement
is found. Charm decays can also provide unique new physics
signatures; the status of present searches is reviewed. Finally,
charm data relevant for improving beauty decay measurements are
presented.}]
\section{Introduction}
The charm quark has played a unique role in particle physics for
more than three decades. Its discovery by itself was an important
validation of the Standard Model, as its mass and most of its
relevant properties were predicted before any experimental
signature for charm was available. Since then, much has been
learned about the properties of charmed hadronic systems.

Experiments operating at the $\psi(3770)$ resonance, near
threshold for $D\bar{D}$ production, such as MARK III at SPEAR,
performed the initial exploration of charm
phenomenology.\cite{markiii} Later, higher energy machines, either
fixed target experiments operating at hadron machines or higher
energy $e^+e^-$ colliders, entered this arena, with much bigger
data samples. In recent years, we have seen a renewed interest in
studying open charm in $e^+e^-$ colliders with a center-of-mass
energy close to $D\bar{D}$ threshold. The CLEO-c
experiment\cite{yellowbook} at CESR, has collected a sample of 281
pb$^{-1}$ at the $\psi(3770)$ center-of-mass energy. This
experiment is poised to accumulate a total integrated luminosity
of the order of 1 fb$^{-1}$ at the $\psi(3770)$ and a similar size
sample at an energy optimal to study $D_S$ decays. The BES-II
experiment, at BEPC, has published results based on 33 pb$^{-1}$
accumulated around the $\psi(3770)$. It has an ongoing upgrade
program both for the detector (BESIII) and the machine (BEPCII),
designed as a charm factory with $10^{33} {\rm cm^{-2}s^{-1}}$
peak luminosity.\cite{bes3}

Several features distinguish charm Its mass ($\cal{O}$(1.5) GeV)
makes it an ideal laboratory to probe QCD in the non-perturbative
domain. In particular, a comparative study of charm and beauty
decays may lead to more precise theoretical predictions for key
quantities necessary for accurate determination of important
Standard Model parameters. On the other hand, once full QCD
calculations have demonstrated control over hadronic
uncertainties, charm data can be used to probe the Yukawa sector
of the Standard model. Finally, charm decays provide a unique
window on new physics affecting the u-type quark dynamics. For
example, it is the only u-type quark that can have flavor
oscillations. Moreover, some specific new physics models predict
enhancements on CP violation phases in D decays, beyond the
$10^{-3}$ level generally predicted within the Standard
Model.\cite{petrov}

The charge-changing transitions involving quarks feature a complex
pattern, that is summarized by a 3$\times$3 unitary matrix, the
Cabibbo-Kobayashi-Maskawa ($CKM$) matrix:
\begin{equation}
V_{CKM} = \left(\begin{array}{ccc}
V_{ud} & V_{us}   & V_{ub} \\
V_{cd} &  V_{cs} & V_{cb} \\
V_{td} &  V_{ts}& V_{tb}
\end{array}\right).
\end{equation}
These 9 complex couplings are described by 4 independent
parameters. In the Wolfenstein approximation,\cite{Wolfenstein}
the CKM matrix is expressed in terms of the four parameters
$\lambda$, $A$, $\rho$, and $\eta$, and is expanded in powers of
$\lambda$:
\begin{equation}
\left(
\begin{array}{ccc}
1-\lambda^2/2 &  \lambda & A\lambda^3(\rho-i\eta) \\
-\lambda &  1-\lambda^2/2&
A\lambda^2 \\
A\lambda^3(1-\rho-i\eta) &  -A\lambda^2& 1
\end{array}
\right).
\end{equation}

The parameters $\lambda$, $A$, $\rho$ and $\eta$ are fundamental
constants of nature, just as basic as $G$, Newton's constant, or
$\alpha_{EM}$.

$B$ meson semileptonic decays (determining $|V_{ub}|/|V_{cb}|$)
and neutral $B$ flavor oscillations provide crucial constraints to
determine the $CKM$ parameters $\rho$ and $\eta$. In both cases,
hadronic matrix elements need to be evaluated to extract these
parameters from the experimental data. Due to the relatively small
masses of the $b$ and $c$ quarks, strong interactions effects are
of a non-perturbative nature. Lattice QCD calculations seem the
ideal approach to tackle this problem. However, a realistic
simulation of quark vacuum polarization has eluded theorists for
several decades, thus limiting lattice QCD results to the
so-called ``quenched approximation." A new unquenched approach,
based on a Symanzik-improved staggered-quark
formalism,\cite{davies} bears the promise of precise predictions
on some key observables.\cite{kronfeld} The main ingredients of
the new approach are: improved staggered quarks representing sea
and valence quarks, chiral perturbation theory for staggered
quarks and heavy quark effective theory (HQET) for the heavy
quarks.\cite{kronfeld} This formalism is expected to deliver
predictions soon on some ``golden" physical quantities with errors
of a few \%. They are matrix elements that involve one hadron in
the initial state and one or no stable hadrons in the final state,
and they require that the chiral perturbation theory is
``well-behaved" for the specific mode under consideration. Several
processes relevant for the study of quark mixing fall in this
category. Important examples include the leptonic decay constants
$f_{B_{(s)}}$ and $f_{D_{(s)}}$ and semileptonic decay form
factors. Checks on theory predictions for key ``golden quantities"
are under way\cite{yellowbook,bes3} and may validate the theory
inputs for the corresponding quantities in beauty decays.

 \section{The decay constant $f_{D^+}$.}

$CKM$ unitarity tests include constraints from
$B^0_{(s)}\bar{B}^0_{(s)}$ oscillations. The theoretical inputs,
 are $\sqrt{\hat{B_d}}f_{B_d}$,
$\sqrt{\hat{B_s}}f_{B_s}$, or $\xi\equiv
{\sqrt{\hat{B_s}}f_{B_s}}/\sqrt{\hat{B_d}}f_{B_d}$, where
$\hat{B}_i$ represents the relevant ``bag parameter", the
correction for the vacuum insertion approximation, and $f_{B_i}$
represents the corresponding decay constant.  It is thus important
to validate the theoretical uncertainties, and a proposed strategy
is to use the corresponding observables in $D$ decays for this
purpose.
 The decay $D^+\to
\ell^+\nu$ proceeds by the $c$ and $\overline{d}$ quarks
annihilating into a virtual $W^+$, with a decay
width\cite{Formula1} given by:
\begin{eqnarray}\label{fd:eq}
\Gamma(D^+\to \ell^+\nu) = {G_F^2\over
8\pi}f_{D^+}^2m_{\ell}^2M_{D^+} \\
\nonumber  \left(1-{m_{\ell}^2\over M_{D^+}^2}\right)^2
\left|V_{cd}\right|^2~~~, \label{eq:equ_rate}
\end{eqnarray}
where $M_{D^+}$ is the $D^+$ mass, $m_{\ell}$ is the mass of the
final state lepton, $|V_{cd}|$ is a CKM matrix element that we
assume to be equal to $|V_{us}|$, and $G_F$ is the Fermi coupling
constant. Due to helicity suppression, the rate goes as
$m_\ell^2$; consequently the electron mode $D^+ \to e^+\nu$ has a
very small rate in the Standard Model. The relative widths are
$2.65:1:2.3\times 10^{-5}$ for the $\tau^+ \nu$, $\mu^+ \nu$ and
$e^+ \nu$ final states, respectively.

\begin{figure}
\epsfxsize160pt\figurebox{120pt}{160pt}{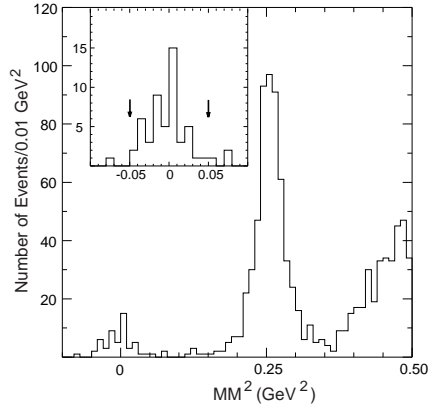}
\caption{CLEO-c $MM^2$ using $D^-$ tags and one opposite charged
track with no extra energetic clusters.$^{10}$ The insert shows
the signal region for $D^+\rightarrow \mu\nu_{\mu}$ enlarged; the
defined signal region is shown between the two arrows.}
\label{fd:data}
\end{figure}

CLEO-c was the first experiment to have a statistically
significant $D^+\rightarrow \mu \nu$ signal,\cite{bonvicinietal}
and has now published an improved measurement of $ f
_{D^+}$.\cite{new-fd-cleo} They use a tagging technique similar to
that developed by the MARK III collaboration,\cite{markiii} where
one $D$ meson is reconstructed in a low background hadronic
channel and the remaining tracks and showers are used to study a
specific decay mode. The relatively high single tag yield makes
this technique extremely useful.\footnote{Throughout this paper
charge conjugate particles are implied unless specifically noted.}
They reconstruct the $D^-$ meson in one of six different decay
modes and search for $D^+\rightarrow \mu \nu$ in the rest of the
event. The existence of the neutrino is inferred by requiring the
missing mass squared ($MM^2$) to be consistent with zero. Here:
\begin{equation}
MM^2= (E_{beam}-E_{\mu^+})^2-(-\vec{p}_{D^-}-\vec{p}_{\mu
^+})^2,\nonumber
\end{equation}
where $\vec{p}_{D^-}$ is the three-momentum of the fully
reconstructed $D^-$. Events with additional charged tracks
originating from the event vertex or unmatched energy clusters in
the calorimeters with energy greater than 0.250 GeV are vetoed.
These cuts are very effective in reducing backgrounds.
Efficiencies are mostly determined using data, while backgrounds
are evaluated either with large Monte Carlo samples or with data.
Fig.~\ref{fd:data} shows the measured $MM^2$, with a 50 event peak
in the interval [-0.050 GeV$^2$,+0.050 GeV$^2$], approximately
$\pm 2\sigma$ wide. The background is evaluated as $2.81\pm
0.30\pm 0.27$ events. This implies:
\begin{equation}
{\cal B}(D^+\rightarrow \mu^+\nu_{\mu}) = (4.40 \pm 0.66
^{+0.09}_{-0.12})\times 10^{-4}.\nonumber
\end{equation}

The  decay constant $ f_{D^+}$ is derived from Eq.~\ref{fd:eq}
using $\tau_{D^+}=1.040\pm\ 0.007$ ps,\cite{pdg2004} and
$|V_{cd}|= 0.2238 \pm 0.0029$,\cite{vcd} yielding:
\begin{equation}
 f_{D^+}= (222.6 \pm 16.7 ^{+2.8}_{-3.4})\ {\rm MeV}.
\end{equation}
The same tag sample is used to search for
$D^+\rightarrow e^+ \nu_{e}$. No signal is found, corresponding to
a 90\% cl upper limit ${\cal B}(D^+\rightarrow e^+ \nu _e)<
2.4\times 10^{-5}$. These measurements are much more precise than
previous observations or limits.\cite{prepr-11} The very small
systematic error is achieved through very careful background and
efficiency studies, involving large Monte Carlo and data samples.

\begin{figure*}
\centerline{\epsfxsize=3in\epsffile{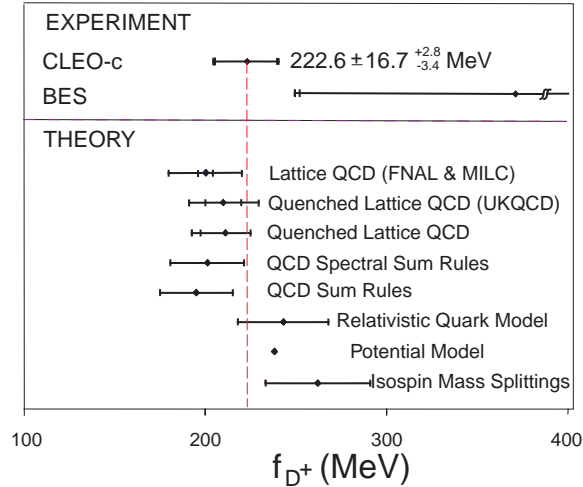}} \caption{Summary
of theoretical predictions and experimental data for
$f_{D^+}$.}\label{fig:fd}
\end{figure*}

Fig.~\ref{fig:fd} summarizes the present experimental
data\cite{new-fd-cleo}$^,$\cite{prepr-11} and the various
theoretical predictions for the decay
constant.\cite{Lat:Taiwan}$^-$\cite{Isospin} The latest lattice
QCD result, performed by the Fermilab lattice, MILC and HPQCD
collaborations, working together,\cite{new-fd-lattice} is the
first to include three quark flavors fully unquenched  and was
published shortly before the \hbox{CLEO-c} updated result. It is
consistent with the CLEO-c result with a 37\% confidence level.

\section{Semileptonic decays}
The study of $D$ meson semileptonic decays is another important
area of investigation. In principle, charm meson semileptonic
decays provide the simplest way to determine the magnitude of
quark mixing parameters: the charm sector allows direct access to
$|V_{cs}|$ and $|V_{cd}|$. Semileptonic decay rates are related to
$|V_{cx}|^2$ via matrix elements that describe strong interactions
effects. Traditionally, these hadronic matrix elements have been
described in terms of form factors cast as a function of the
Lorentz invariant $q^2$, the invariant mass of the electron-$\nu$
pair. Experimental determinations of these form factors are
performed through the study of the differential decay width
$d\Gamma /dq^2$.

\subsection{Goals in semileptonic decays}

If we assume that $V_{cs}$ and $V_{cd}$ are known, experiments can
determine the form factor shape as well as their normalization.
Form factors have been evaluated at specific $q^2$ points in a
variety of phenomenological models,\cite{stone:sl} where the shape
is typically assumed. More recently, lattice QCD
calculations\cite{aubinetal} have predicted both the normalization
and shape of the form factors in $D\rightarrow K\ell\nu$ and
$D\rightarrow \pi\ell\nu$. Note, that we can form ratios between
leptonic and exclusive semileptonic branching fractions that can
provide direct theory checks without any $CKM$ input.

On the other hand, if we use validated theoretical results as
inputs, we can derive direct measurements for $V_{cs}$ and
$V_{cd}$; the most accurate determinations of these parameters
presently require some additional input information, such as
unitarity. Thus we could extend the unitarity checks of the CKM
matrix beyond the first row.

The study of charm semileptonic decays may contribute to a precise
determination of the CKM parameter $|V_{ub}|$. A variety of
theoretical approaches have been proposed  to use constraints
provided by charm decays to reduce the model dependence in the
extraction of $|V_{ub}|$ from exclusive charmless $B$ semileptonic
decays. In particular, if HQET is applicable both to the $c$ and
$b$ quarks, there is a SU(2) flavor symmetry that relates the form
factors in $D$ and $B$ semileptonic decays.\cite{isgur-wise} For
example, a flavor symmetry relates the form factors in
$D\rightarrow \pi \ell \nu$ are related to the ones in
$B\rightarrow \pi \ell \bar{\nu}$, at the same $E\equiv{\bf
v}\cdot {\bf p}_{\pi}$, where ${\bf v}$ is the heavy meson
4-velocity and ${\bf p}_{\pi}$ is the $\pi$ 4-momentum. The
original method has been further refined;\cite{ben} the large
statistics needed to implement these methods may be available in
the near future.

\begin{table*}[hbt] \caption{Summary of recent absolute branching
fraction measurements of exclusive $D^+$ and $D^0$ semileptonic
decays.}\label{tab:brsemil}\begin{center}
\begin{tabular}{|l|c|c|c|}\hline
Decay mode & {\cal B}(\%) [CLEO-c]\cite{cleoc:excl} & {\cal B}(\%)
[BES]\cite{bes:excl} & {\cal B}(\%)
[average]$^\aleph$\\
\hline $D^0\rightarrow K^- e^+ \nu _e$ & $3.44\pm 0.10 \pm 0.10$ &
$3.82 \pm 0.40 \pm 0.27$  & $3.54 \pm 0.11$ \\ \hline
$D^0\rightarrow \pi^- e^+ \nu _e$ & $0.262\pm 0.025 \pm 0.008$ &
$0.33 \pm 0.13\pm 0.03$ & $0.285 \pm 0.018$ \\ \hline
$D^0\rightarrow K^{\star -} e^+ \nu _e$ & $2.16\pm 0.15 \pm 0.08$
& ~~ & $2.14 \pm 0.16$ \\ \hline $D^0\rightarrow \rho^- e^+ \nu
_e$ & $0.194\pm 0.039\pm 0.013$ & ~~ & $0.194\pm 0.039\pm 0.013$
\\ \hline
\hline $D^+\rightarrow \bar{K}^0 e^+\nu _e$ & $8.71 \pm 0.38 \pm
0.37$ & ~~ & $8.31\pm 0.44$ \\ \hline $D^+\rightarrow \pi^0
e^+\nu _e$ & $0.44 \pm 0.06 \pm 0.0.03$ & ~~ & $0.43 \pm 0.06$ \\
\hline $D^+\rightarrow \bar{K}^{\star 0} e^+\nu _e$ & $5.56 \pm
0.27 \pm 0.23$ & ~~ & $5.61\pm 0.32$ \\ \hline $D^+\rightarrow
\rho^0
e^+\nu _e$ & $0.21 \pm 0.04 \pm 0.01$ & ~~ & $0.22\pm 0.04$ \\
\hline $D^+\rightarrow \omega e^+\nu _e$ & $0.16 ^{+0.07}_{-0.06}
\pm 0.01$ & ~~ & $0.16 ^{+0.07}_{-0.06}$\\ \hline
\end{tabular}\break\end{center}
$^{\aleph}${\small The averages reported here include all the
branching fractions reported in the PDG 2004 for $D\rightarrow X
e^+\nu _{e}$ and the CLEO-c and BES-II data. Indirect measurements
are normalized with respect to the hadronic\cite{He:05} and
average semileptonic branching ratios included in this report.}
\end{table*}
\subsection{Semileptonic branching
fractions: the data}

\begin{figure}[hbt]
\epsfxsize170pt \figurebox{120pt}{160pt}{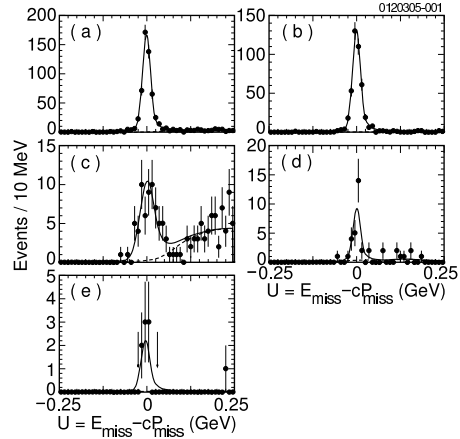}
\caption{Fits (solid lines) to the $U$ distributions in
CLEO-c$^{28}$ data (dots with error bars) for the five $D^+$
semileptonic modes: (a) $D^+\rightarrow \bar{K}^0e^+\nu_e$,
(b)$D^+\rightarrow \bar{K}^{\star 0}e^+\nu_e$, (c) $D^+\rightarrow
\pi^0 e^+\nu_e$, (d)$D^+\rightarrow \rho^0 e^+\nu_e$,
(e)$D^+\rightarrow \omega e^+\nu_e$. The arrows in (e) show the
signal region. The background (in dashed lines) is visible only in
(c) and (d).}\label{dplus-sl}
\end{figure}

BES-II\cite{bes:excl} and CLEO-c\cite{cleoc:excl} have recently
presented data on exclusive semileptonic branching fractions.
BES-II results are based on 33 pb$^{-1}$; CLEO-c's results are
based on the first 57 pb$^{-1}$ data set. Both experiments use
tagged samples and select a specific final state through the
kinematic variable:
\begin{equation}
U\equiv E_{miss} -|c\vec{p}_{miss}|,\nonumber
\end{equation}
where $E_{miss}$ represents the missing energy and $\vec{p}$
represents the missing momentum of the $D$ meson decaying
semileptonically. For signal events, $U$ is expected to be 0,
while other semileptonic decays peak in different regions.
Fig.~\ref{dplus-sl} shows the $U$ distribution for 5 exclusive
$D^+$ decay modes reported by CLEO-c, which demonstrate that $U$
resolution is excellent, thus allowing  a full separation between
Cabibbo suppressed and Cabibbo favored modes. Table
\ref{tab:brsemil} summarizes the recent measurements from CLEO-c
and BES-II, as well world averages obtained from the results
presented in this paper and the previous measurements of ${\cal
B}(D\rightarrow X_i e^+\nu _e)$ reported in the PDG
2004.\cite{pdg2004}

CLEO-c uses the two tagging modes with lowest background
($\bar{D}^0\rightarrow K^+\pi^-$ and $D^-\rightarrow
K^+\pi^-\pi^-$) to measure the inclusive $D^0$ and $D^+$
semileptonic branching fractions.\cite{cleoc:incl} Table
\ref{dincl} summarizes the measured semileptonic branching
fractions, and it also includes the sum of the branching fractions
for $D$ decay into all the known exclusive modes. The CLEO-c data
have been used in this comparison, as they dominate the present
world average: the exclusive modes are consistent with saturating
the inclusive semileptonic branching fraction at a 41\% confidence
level in the case of the $D^+$ and 18\% confidence level in the
case of the $D^0$.

The preliminary inclusive branching fractions can be translated
into  inclusive semileptonic widths $\Gamma^{sl}_{D^+}$ and
$\Gamma^{sl} _{D^0}$, using the known $D$
lifetimes,\cite{pdg2004}. These widths are expected to be equal,
modulo isospin violations, and indeed the measured ratio
$\Gamma^{sl}_{D^+}/\Gamma^{sl} _{D^0} = 1.01 \pm 0.03 \pm 0.03$:
thus isospin violations are limited to be below $\sim$ 4\%.

\begin{table}[hbt]
\caption{Comparison between exclusive$^{28}$ and preliminary
inclusive$^{29}$ results from CLEO-c.}\label{dincl}
\begin{tabular}{|l|c|}
\hline Mode & ${\cal B}$ (\%)
\\\hline
$(D^0 \rightarrow X \ell \nu_e)$ & $6.45 \pm 0.17 \pm 0.15$\\
\hline
 $\Sigma _i {\cal B}((D^0\rightarrow X_i \ell \nu_e)$&
 $6.1\pm 0.2\pm 0.2$ \\\hline\hline
 $(D^+ \rightarrow X \ell \nu_e)$ & $16.19 \pm 0.20 \pm 0.36$\\
\hline
 $\Sigma _i {\cal B}((D^+\rightarrow X_i \ell \nu_e)$ &
 $15.1\pm 0.50 \pm 0.50$\\\hline

\end{tabular}
\end{table}

\subsection{Form factors for $D\rightarrow K(\pi) \ell \nu$}

 Recently, non-quenched lattice $QCD$ calculations for
$D\rightarrow K\ell \bar{\nu}$ and $D\rightarrow \pi \ell \nu$
have been reported.\cite{aubinetal} The chiral extrapolation is
performed at fixed $E =\vec{v}\cdot\vec{p}_P$, where $E$ is the
energy of the light meson in the center-of-mass $D$ frame,
$\vec{v}$ is the unit 4-velocity of the $D$ meson, and $\vec{p}_P$
is the 4-momentum of the light hadron $P$ ($K$ or $\pi$). The
results are presented in terms of a parametrization originally
proposed by Becirevic and Kaidalov ($BK$):\cite{Becirevic:1999kt}
\begin{eqnarray}\label{eq:BK}
f_+(q^2) = \frac{F}{(1-\tilde{q}^2)(1-\alpha\tilde{q}^2)},\\
\nonumber
 f_0(q^2) = \frac{F}{1-\tilde{q}^2/\beta},
\end{eqnarray}
where $q^2$ is the 4-momentum of the electron-$\nu$ pair,
$\tilde{q}^2=q^2/m_{D_x^{*}}^2$, and $F=f_+(0)$, $\alpha$ and
$\beta$ are fit parameters. This formalism models the effects of
higher mass resonances other than the dominant spectroscopic pole
($D^{\star +}_S$ for the $K\ell \nu$ final state and $D^{\star +}$
for $\pi\ell \nu$).\cite{fajfer}

The form factors $f_+(q^2)$ govern the corresponding semileptonic
decays. The lattice QCD calculation obtains the parameters shown
in Table~\ref{tab:results}.
\begin{table}[t]
\caption{Fit parameters in Eq.~(8), decay rates and CKM matrix
elements. The first errors are statistical; the second
systematic.$^{24}$} \label{tab:results}
\begin{tabular}{|l|c|c|c|}
\hline $P$ & $F$& $\alpha$ & $\beta$ \\ \hline $\pi$ &0.64(3)(6)
&0.44(4)(7) &1.41(6) (13) \\\hline
 $K$  &0.73(3)(7) &0.50(4)(7) &1.31(7)(13) \\
\hline
\end{tabular}
\end{table}

\begin{figure}
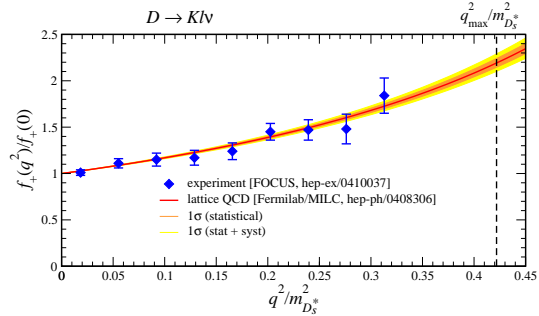

\epsfxsize200pt \figurebox{120pt}{160pt}{f+shape.eps}
\caption{Shape of the form factor for $D\rightarrow K\ell\nu$:$^7$
MILC-Fermilab calculation compared with the non parametric data
from FOCUS.}\label{shape}
\end{figure}

The FOCUS experiment\cite{focus:ff} performed a non-parametric
measurement of the shape of the form factor in $D\rightarrow K \mu
\nu _{\mu}$. Fig.~\ref{shape} shows the lattice QCD predictions
for $D\rightarrow K\ell \nu$ with the FOCUS data points
superimposed. In addition, they studied the shape of the form
factors $f_+(q^2)$ for $D\rightarrow K\mu\nu _{\mu}$ and
$D\rightarrow \pi\mu\nu _{\mu}$ with two different fitting
functions: the single pole, traditionally used because of the
conventional ansatz of several quark models,\cite{stone:sl} and
the $BK$ parametrization discussed before. Table \ref{fit-data}
shows the fit results obtained from FOCUS and CLEO
III,\cite{cleo3-pilnu} compared to the lattice QCD predictions.
Both experiments obtain very good fits also with simple pole form
factors, however the simple pole fit does not yield the expected
spectroscopic mass. For example, FOCUS obtains
$m_{pole}(D^0\rightarrow K\mu \nu_{\mu})=(1.93\pm 0.05 \pm 0.03)$
GeV/c$^2$ and $m_{pole}(D^0\rightarrow \pi\mu \nu
_{\mu})=(1.91^{+0.30}_{-0.15} \pm 0.07$) GeV/c$^2$, while the
spectroscopic poles are, respectively, $2.1121\pm 0.0007$
GeV/c$^2$ and $2.010\pm 0.0005$. This may hint that other higher
order resonances are contributing to the form
factors.\cite{fajfer}
 It has been argued,\cite{hill} that even
the $BK$ parametrization is too simple and that a three parameter
form factor is more appropriate. However, this issue can be
resolved only by much larger data samples, with better sensitivity
to the curvature of the form factor near the high recoil region.

\begin{table}
\caption{Measured shape parameter $\alpha$ compared to lattice QCD
predictions.}\label{fit-data}
\begin{center}
\begin{tabular}{|c|c|}\hline
\multicolumn{2}{|c|}{$\alpha(D^0\rightarrow K\ell \nu)$}\\\hline
lattice QCD\cite{aubinetal}& $0.5\pm 0.04\pm 0.07$ \\\hline
FOCUS\cite{focus:ff} & $0.28\pm 0.08\pm 0.07$\\\hline
CLEOIII\cite{cleo3-pilnu} & $0.36\pm 0.10^{+0.03}_{-0.07}$\\\hline
Belle\cite{belle:ff}& $0.40\pm 0.12\pm 0.09 $\\ \hline
\multicolumn{2}{|c|}{$\alpha(D^0\rightarrow \pi\ell \nu$)}\\\hline
lattice QCD\cite{aubinetal}& $0.44\pm 0.04\pm 0.07$
\\\hline CLEOIII\cite{cleo3-pilnu} & $0.37^{+0.20}_{-0.31}\pm
0.15$\\\hline Belle\cite{belle:ff}& $0.03\pm 0.27 \pm 0.13$
\\\hline
\end{tabular}
\end{center}
\end{table}

By combining the information of the measured leptonic and
semileptonic width, a ratio independent of $|V_{cd}|$ can be
evaluated: this is a pure check of the theory. We evaluate the
ratio $R\equiv\sqrt{\Gamma(D^+\rightarrow \mu \nu
_{\mu})/\Gamma(D\rightarrow \pi e^+\nu _e)}$. We assume isospin
symmetry, and thus $\Gamma(D\rightarrow \pi e^+\nu
_e)=\Gamma(D^0\rightarrow \pi^- e^+\nu _e)=2\Gamma(D^+\rightarrow
\pi^0 e^+\nu _e)$. For the theoretical inputs, we use the recent
unquenched lattice QCD calculations in three
flavors,\cite{new-fd-lattice,aubinetal} as they reflect the state
of the art of the theory and have been evaluated in a consistent
manner. The result is:
\begin{equation}
R^{th}_{sl}=\sqrt{\frac{\Gamma^{th}(D^+\rightarrow \mu\nu
_{\mu})}{\Gamma^{th}(D\rightarrow \pi e\nu _{e})}}=0.212\pm
0.028,\nonumber
\end{equation}
The quoted error is evaluated through a careful study of the
theory statistical and systematic uncertainties, assuming Gaussian
errors. The corresponding experimental quantity is calculated
using the CLEO-c $f_D$ and isospin averaged $\Gamma (D\rightarrow
\pi e^+\nu_e)$; we obtain:
\begin{equation}
R^{exp}_{sl}=\sqrt{ \frac {\Gamma^{exp}(D^+\rightarrow\mu\nu)}
{\Gamma^{exp}(D\rightarrow\pi e \nu _e)}}= 0.249\pm
0.022.\nonumber
\end{equation}
The theory and data are consistent at 28\% confidence level, that
represents a good agreement.

\section{The $CKM$ Matrix}

An important goal of the next generation of precision experiments
is to perform direct measurements of each individual parameter.
This will enable us to perform additional unitarity checks with
precision similar to the one achieved now with the first
row.\cite{vcd} In particular, $V_{cd}$ and $V_{cs}$ are now
determined with high precision, but using unitarity
constraints.\cite{pdg2004} The most recent results from LEP II,
using the $W\rightarrow \ell \nu$ branching fraction, and
additional inputs from other $CKM$ parameter measurement is
$V_{cs}=0.976\pm 0.014$.\cite{lep-win05} The unitarity constraint
implies $V_{cd}\sim V_{us}= 0.2227\pm 0.0017$.\cite{vcd}

If we use the theoretical form factors as inputs, we can extract
$|V_{cs}|$ and $|V_{cd}|$ from the branching fractions reported in
this paper. The results, obtained using the form factors from the
unquenched lattice QCD calculation\cite{aubinetal} and the isospin
averaged semileptonic widths from \hbox{CLEO-c}\cite{cleoc:excl}
are:
\begin{equation}
|V_{cs}|=  0.957\pm 0.017(exp)\pm 0.093(th)\nonumber
\end{equation}
\begin{equation} |V_{cd}| = 0.213 \pm 0.008(exp)\pm
0.021(th)\nonumber
\end{equation}
A unitarity check derived uniquely from these measurements yields:
\begin{equation}
1-|V_{cs}|^2+|V_{cd}|^2+|V_{cb}|^2 = 0.037\pm 0.181
(tot).\nonumber
\end{equation}
The mean $V_{ci}$ and their errors have been derived from careful
application of the theoretical quantities and their stated
statistical and systematic errors.

These determinations are not yet competitive, but it will be
interesting to see the results of future estimates, when the
accuracy is comparable to the one achieved in the first row.
\section{Charm as a probe for New Physics}
The study of charm decays provides a unique opportunity for
indirect searches for physics beyond the Standard Model. In
several dynamical models, the effects of new particles observed in
$c$, $s$ and $b$ transitions are
correlated.\cite{petrov}$^,$\cite{nir} Possible new physics
manifestations involve three different facets: $D^0\bar{D}^0$
oscillations, $CP$ violation and rare decays.

\subsection{{$D^0\bar{D}^0$} oscillations}

Two main processes contribute to $D^0\bar{D}^0$ oscillations. The
short distance physics effects are depicted by higher order
Feynman  diagrams, such box or loop diagrams that influence the
mass difference $\Delta M$. These diagrams are sensitive to new
physics, through the interference with contributions with similar
topology including exotic particles in place of the $d,\ s,\ b$
quarks present in the Standard Model loop. In addition, there is a
coupling between $D^0$ and $\bar{D}^0$ induced by common final
states such as $K\bar{K}$, $\pi\pi$ and $K\pi$. As the
intermediate states are real, one conjectures that only the
difference in lifetime $\Delta \Gamma$ is affected by this
coupling. Thus, $\Delta\Gamma$ is expected to be dominated by
Standard Model processes.

$D^0\bar{D}^0$ mixing haw been studied with a variety of different
experimental methods, several of which suffer from a variety of
additional complications.

The first approach, which has been pursued by a variety of
experiments,\cite{kpi-cleo}$^-$\cite{kpi-babar} is the study of
the ``wrong-sign" hadronic decays such as $D^0\rightarrow
K^+\pi^-$. These decays occur via two paths: oscillation of $D^0$
into $\bar{D}^0$, followed by the Cabibbo favored
$\bar{D}^0\rightarrow K^+\pi^-$, or doubly Cabibbo suppressed
decays $D^0\rightarrow K^+\pi^-$. The two channels interfere and
thus there is an additional parameter that affects the wrong-sign
rate: the strong phase $\delta$ between $D^0\rightarrow K^+\pi^-\
{\rm and}\ K^-\pi^+$ decays. Moreover it has been
argued\cite{Wolfenstein:dmix} that $CP$ violation may have non
negligible effects too. Thus experiments typically perform a
variety of fits for the modified variables $x^\prime \equiv
x\cos{\delta}+y \sin{\delta}$ and $y^\prime \equiv -x\sin{\delta}
+ y \cos{\delta}$, under different CP violation assumptions. Table
\ref{dmixkpi} summarizes the results of the most generic fit,
allowing for a CP violating term.

\begin{table}
\caption{$D^0\rightarrow K^+\pi^-$ analysis. Only results of the
fits allowing for $CP$ violation are included.}\label{dmixkpi}
\begin{center}
\begin{tabular}{|l|c|}
\hline Experiment & Fit Result ($\times 10^3$)\\\hline
CLEO\cite{kpi-cleo} & $0< x^{\prime 2} < 0.82 $\\
CLEO\cite{kpi-cleo} & $-58 < y^{\prime } < 10 $\\\hline
FOCUS\cite{kpi-focus}&$0< x^{\prime 2}< 8 $\\
FOCUS\cite{kpi-focus}&$-112 <y^{\prime }< 67$ -\\\hline
Belle\cite{kpi-belle}&$0< x^{\prime 2} < 0.89$\\
Belle\cite{kpi-belle}&$-30<y^\prime <27$\\\hline
BaBar\cite{kpi-babar}& $0< x^{\prime} < 2.2$\\
BaBar\cite{kpi-babar}& $-56<y^\prime <39$\\\hline
\end{tabular}
\end{center}
\end{table}

A second class of measurements involves the study of $y_{CP}$:
namely the normalized lifetime difference of $D^0\bar{D}^0$ $CP$
eigenstates. In presence of $CP$ violation, $y_{CP}$ is a linear
combination of $x$ and $y$ involving the $CP$ violation phase
$\phi$. Table~\ref{sum-ycp} summarizes experimental data on
$y_{CP}$. The average is positive, although still consistent with
0.

\begin{table}
\caption{Summary of $y_{CP}$ results.}\label{sum-ycp}
\begin{center}
\begin{tabular}{|l|c|}
\hline Experiment & $y_{CP}$(\%)\\\hline FOCUS\cite{ycp-focus} &
$3.4\pm 1.4 \pm 0.7$\\\hline CLEO\cite{ycp-cleo} &  $-1.2\pm
2.5\pm 1.4$\\\hline Belle, untagged\cite{belle-untg}&$-0.5\pm
1.0\pm0.8$\\\hline Belle, tagged\cite{belle-tag}&$1.2\pm 0.7\pm
0.4$\\\hline BaBar\cite{ycp-babar}& $0.8\pm 0.4
^{+0.5}_{-0.4}$\\\hline
\end{tabular}
\end{center}
\end{table}

The study of semileptonic $D$ decays allows the determination of
another combination of mixing parameters. Experiments study the
ratio $r_M$ defined as:
\begin{equation}
r_M = \frac{\int_0^\infty {\cal P}(D^0\rightarrow
\bar{D}^0\rightarrow X^+\ell \bar{\nu})}{\int_0^\infty {\cal
P}(D^0\rightarrow X^+\ell \bar{\nu})}\approx
\frac{x^2+y^2}{2}.\nonumber
\end{equation} Table~\ref{dmix:sl} summarizes the sensitivity
achieved by present experiments to $r_M$.

\begin{table}
\caption{Summary of mixing limits (95 \% cl) from $D^0$
semileptonic decay studies.}\label{dmix:sl}
\begin{center}
\begin{tabular}{|l|c|c|}
\hline Experiment & $R_M$& $\sqrt{x^2+y^2}$\\\hline
CLEO\cite{sl-cleo} & 0.0091 & 0.135 \\\hline BaBar\cite{sl-babar}
& 0.0046 & 0.1
\\\hline Belle\cite{sl-belle} & 0.0016 & 0.056 \\\hline
\end{tabular}
\end{center}
\end{table}

Finally, a very interesting analysis method has been implemented
by the CLEO experiment: they have studied the channel
$D^0\rightarrow K_S^0\pi^+\pi^-$. Cabibbo favored final states,
such as $K^{\star -}\pi ^+$, and doubly-Cabibbo suppressed
channels, such as $K^{\star +}\pi ^-$ interfere. They generalize
the methodology that they used to identify the resonance
substructure of this decay\cite{cleo-ks2pia} to the case where the
time-dependent state is a mixture of $D^0$ and
$\bar{D}^0$.\cite{cleo-ks2pib}  In this case, the parameters $x$
and $y$ affect the time-dependent evolution of this system. This
time-dependent Dalitz plot analysis can be used to extract the
mixing and $CP$ violation parameters. They obtain $(-4.5<x<9.3)$\%
and $(-6.4<y<3.6)$\%, It is interesting to note that this
constraint has sensitivity comparable to other limits obtained
from a much larger data sample.

\subsection{$CP$ violation}
Within the Standard Model, $CP$ violation effects in $D$ decays
are expected to be negligible small, as they are introduced by box
diagrams or penguin diagrams containing a virtual $b$ quark: thus
they involve a strong $CKM$ suppression ($V_{cb}V^{\star}_{ub})$.
In contrast with the $D^0\bar{D}^0$ mixing case, where the vast
theoretical effort  devoted to pin down the Standard Model
predictions did not yield a clear-cut result, there is a wide
consensus that observing $CP$ violation in $D$ decays at a level
much higher than ${\cal O}(10^{-3})$ will constitute an
unambiguous signal of new physics. There is a vast array of
studies that can be undertaken:\cite{petrov} exploring $CP$
violation effects on mixing observables, searching for direct $CP$
violation effects in $D^0$, $D^+$ and $D^+_S$ decays and, finally,
studies of $D\bar{D}$ pairs near threshold, that exploit the
quantum coherence of these states.

In general, experimental sensitivity is ${\cal
O}(1)$\%.\cite{petrov} Recent results from BaBar,\cite{babar-cpv}
Belle\cite{belle-cpv}, and CLEO\cite{cleo-cpv} have explored $CP$
violation in 3-body $D$ decays. Babar obtains ${\cal
A}(D^+\rightarrow K^-K^+\pi^+)= (1.4\pm 1.0 \pm 0.8)\%$.  CLEO
obtains  ${\cal A}(D^0\rightarrow \pi^+\pi^-\pi^0)=(1\pm 8
^{+9}_{-7})\%$. Belle obtains ${\cal A}(D^0\rightarrow
K^+\pi^-\pi^0)=(-0.6\pm 5.3\%$ and ${\cal A}(D^0\rightarrow
K^+\pi^-\pi^+\pi^-)=(-1.8\pm 4.4\%$.

A complementary approach involves the study of observables that
are sensitive to $T$ violation,\cite{Ikaros:tviol} such as triple
product correlations in 4-body decays of $D^0$ and $D^+$. This
technique has been pioneered by FOCUS,\cite{FOCUS:leppho} through
the study of triple product correlations in $D^0\rightarrow
K^+K^-\pi^+\pi^-$, $D^+\rightarrow K^0_sK^-\pi^+\pi^-$,
$D^0_S\rightarrow K^0_S K^-\pi^+\pi^-$. Their present sensitivity
is at the level of several percent, dominated by the statistical
error. A significant improvement in the sensitivity of this
technique is expected in future measurements.

\section{Charm as a facet of beauty}
The study of $b$ decays has been one of our richest sources of
information about the Standard Model, as well as a very powerful
constraint on new physics.

As the dominant tree level diagram includes the $b\rightarrow c$
transition, the precision of our  knowledge of the $D$ decay
phenomenology affects quantities associated with $B$ decays in a
variety of ways. For example, the accuracy of the determination of
$D$ hadronic branching fractions has an obvious impact on the
absolute determination of $B$ hadronic branching fractions.
Moreover, the study of specific $CP$ violation observables can be
made more precise through ancillary information coming from $D$
decays. Finally, a precise knowledge of the particle yields in $D$
decays, allow a more precise modelling of inclusive $B$ decays.

\subsection{D absolute branching fractions}
Absolute measurements of $D$ meson branching fractions affect our
knowledge of several many $D$ and $B$ meson decays, from which
$CKM$ parameters are extracted.

CLEO-c has employed tagged samples to obtain new values for the
branching fractions $D^0\rightarrow K^-\pi^+$,  $D^+\rightarrow
K^-\pi^+\pi^+$, and other modes.\cite{He:05}  This powerful
technique, combined with careful efficiency studies based on data,
resulted in an accuracy comparable to the one of present world
averages. They obtain:
$${\cal B}(D^0\rightarrow K^-\pi^+) =(3.91 \pm 0.08 \pm 0.09)\%,$$
and
$${\cal B}(D^+\rightarrow K^-\pi^+\pi^+) =(9.5 \pm 0.2 \pm 0.3)\%$$
Corrections for final state radiation are included in these
branching fractions.

\subsection{$D\rightarrow K_S\pi^+\pi^-$ Dalitz plot analysis and
the determination of the $CKM$ phase $\gamma$.}

The decay $B^\pm \rightarrow DK^\pm$ has been the subject of
intense theoretical effort to devise optimal strategies to measure
the $CKM$ angle $\gamma$. The original proposal by Gronau, London
and Wyler\cite{GLW} uses $D$ decays to $CP$ eigenstates.
Subsequently Atwood, Dunietz and Soni\cite{ads} critiqued this
approach and proposed a method based on $D$ decays to flavor
eigenstates. Finally, there is one method that has received a lot
of attention recently,\cite{fit:dalitz} the extraction of $\gamma$
from a Dalitz plot analysis of $B^{\pm} \rightarrow
D^{(\star)}K^{\pm}\rightarrow K^\pm K_S\pi^+\pi^-$. Charm
factories can help this measurement in a variety of manners: they
can provide information on $D^0\bar{D}^0$ mixing, and measure the
strong phase $\delta$ between the Cabibbo favored and
doubly-Cabibbo suppressed $D^0\rightarrow K^-\pi^+$ and
$\bar{D}^0\rightarrow K^-\pi^+$, and perform unique $D$ Dalitz
plot studies.

The Dalitz plot technique illustrates the contributions that
CLEO-c and, later, BESIII can provide to reduce the uncertainty in
this determination of the angle $\gamma$. This method is
attractive because it involves a $D$ decay with a relatively large
branching fraction. Moreover this three body final state comprises
a very rich resonance substructure, that leads to the expectation
of large strong phases. Recently both BaBar\cite{babar-gamma} and
Belle\cite{belle-gamma} reported measurements on $\gamma\ {\rm
(BaBar)} - \phi _3\ {\rm (Belle)} $ with this method. They obtain:
\begin{eqnarray}
\phi _3= 77^{\circ +17}_{\;\,-19}(\rm stat) \pm{13^\circ}(\rm
sys)\pm{11^\circ}(\rm mod),\cr \nonumber
\gamma= 70^\circ \pm 26^{\circ}(\rm stat) \pm{10^\circ}(\rm
sys)\pm{10^\circ}(\rm mod). \nonumber
\end{eqnarray}
In both cases, the error labeled ``mod," refers to uncertainties
on the resonance substructure of the $K_S\pi^+\pi^-$ Dalitz plot.
Both collaborations find that to achieve a good fit they need to
include two ad-hoc $\pi\pi$ s-wave resonances that describe about
10\% of the data. The study of  $CP$ tagged Dalitz
plots\cite{ansner-ckm05} allows a model dependent determination of
the $D^0$ and $\bar{D}^0$ phase across the Dalitz plot. Using data
samples where the $CP$ eigenstate $(\cal S_{\pm})$ of the $D$ can
be tagged, CLEO-c  is studying the Dalitz plots $\cal S_-
K_S\pi^+\pi^-$, $\cal S_- K_S\pi^+\pi^-$, as well as flavor tagged
$K_S \pi^+\pi^-$ Dalitz plots. A simultaneous fit to these three
Dalitz plots can validate Dalitz plot models and reduce the model
dependence of these results significantly. Alternatively, a model
independent result can be obtained from a binned analysis of the
three $CP$ or flavor tagged Dalitz plot. This work is under
way\cite{rosner-beauty05} and should eventually reduce the model
dependence to a couple of degrees.

\section{Conclusions}

Charm decays provide a rich phenomenology for a variety of
important studies that improve our knowledge of several facets of
the Standard Model, and probe for signatures of new physics.

The experimental study of beauty and charm decays is prospering
through vibrant experimental activity taking place in several
ongoing experiments. The next few years will see an opening up of
our vistas on these decays with the upcoming turn on of LHC and of
a dedicated charm and beauty experiment at a hadron collider,
LHCb. This experiment bears the promise of precision studies that
are poised to explore thoroughly all the possible new physics
manifestations alluded to in this paper.
\section*{Acknowledgments}
I would like to thank the organizers for their tremendous effort
that lead to a very enjoyable and productive conference. I would
also like to acknowledge interesting discussions and scientific
input from D. Asner, A. Kronfeld, M. Okamoto, and S. Stone. This
work was supported by the United States National Science
Foundation.

\end{document}